\documentclass{basi}
\usepackage{txfonts}
\usepackage{graphicx}

\newcommand{\msun}{\ifmmode{\rm M_{\odot}}\else{M$_{\odot}$}\fi}
\newcommand{\arcsec}{\ifmmode{\rm ^{\prime \prime}}\else{$^{\prime
\prime}$}\fi}
\newcommand{\lsun}{\ifmmode{\rm L_{\odot}}\else{L$_{\odot}$}\fi}
\newcommand{\degree}{\ifmmode{\rm ^{\circ}}\else{$^{\circ}$}\fi}
\newcommand\arcdeg{\mbox{$^\circ$}}%
\newcommand{\mic}{\ifmmode{\rm \mu m}\else{$\mu$m}\fi}
\newcommand{\htw}{\ifmmode{\rm H_2}\else{H$_2$}\fi}
\newcommand{\htwo}{\ifmmode{\rm H_2O}\else{H$_2$O}\fi}

\begin{document}
\title[Monte Carlo radiative transfer]{Monte Carlo radiative transfer}
\author[Barbara~A.~Whitney]
       {B.~A.~Whitney,$^{1,2}$\thanks{e-mail:bwhitney@astro.wisc.edu}
        \\
       $^1$Astronomy Department, University of Wisconsin-Madison, 475 N. Charter St., Madison, WI 53706, USA
        \\
       $^2$Space Science Institute, 4750 Walnut Street, Suite 205, Boulder, Colorado 80301, USA}

\pubyear{2011}
\volume{39}
\pagerange{\pageref{firstpage}--\pageref{lastpage}}
\date{Received 2011 April 04; accepted 2011 April 14}

\maketitle
\label{firstpage}

\begin{abstract}
I outline methods for calculating the solution of Monte Carlo Radiative Transfer (MCRT) in   scattering, absorption and emission processes of dust and gas, including polarization.
I provide a bibliography of relevant papers on methods with
astrophysical applications.
\end{abstract}


\section{Introduction}\label{s:intro}

The Monte Carlo method was invented by Stanislaw Ulam and John von Neumann to study neutron transport during 
the atomic bomb program of World War II.  According to Wikipedia, because the work was secret, a code name 
was needed, so they chose Monte Carlo, after the famous Casino in Monaco which Ulam's uncle frequented.
At this time and for several decades after, the pressing radiative transfer problems in astrophysics were in stellar 
atmospheres and interiors, which fortunately are 1-D
problems that could be solved with other, much faster methods.  
Many clever integral and differential equation techniques were devised to calculate sophisticated stellar atmosphere 
models, including line transfer and stellar winds.  These methods
are reviewed in several standard texts, e.g., Mihalas (1978).
Scattering and polarization were always the most complicated aspects of these methods, and were
therefore often ignored.
Not surprisingly, these were tackled very early by S. Chandrasekhar (1946, 1960). 

As radiative transfer began to be applied to other kinds of objects that are not as spherical as stars,
it became necessary to consider multi-dimensional geometries and scattering.  As an example, both forming
and evolved stars are often surrounded by dusty disks and/or clumpy envelopes and outflows.  The asymmetric circumstellar
geometries produce very different spectral energy distributions (SEDs) than 1-D models can account for.
Galaxies can appear bluer than expected if scattering from interstellar dust is not taken into account.
A method that is ideally suited to solve these types of problems is the Monte Carlo method.
I was fortunate to have my thesis advisor, Art Code, suggest this method to study polarization in magnetic
white dwarfs, back in the 1980s.
I then applied this method in the area of star formation, where 2-D radiative transfer proved
very useful in interpreting the disk and bipolar structures of Young Stellar Objects (YSOs).
Since this time, many scientists have developed new methods to calculate, e.g.,
the radiative equilibrium solution for dust,  gas line
and continuum transfer, photoionization, polarization, and relativistic radiative transfer
(references for these methods
and applications are given later in the text).  Now the Monte Carlo method
is in widespread use in astronomy and is an exciting area to get into.

This article is designed for readers who are interested in learning the Monte Carlo method for radiative 
transfer in astrophysics.  It starts with 
the basics needed to write a complete but simple Monte Carlo scattering code (Section 2), and then shows more 
complicated but common scattering problems (Section 3), 
dust emission (Section 4.1-4.5), and gas emission (Section 4.6).  Not everything is described in detail, e.g., line scattering
and gas emission, but numerous references are cited.
I have made an attempt to include the most relevant and up-to-date references
on methods, but I surely have missed some and I apologize for 
this\footnote{Please send me any relevant references and I will update the online version
of this document}.

\section{Monte Carlo basics and a simple scattering problem}

In the Monte Carlo method for radiative transfer (MCRT),  probabilistic methods are used to simulate the transport of individual 
`photon packets' (which we will abbreviate as `photons') through a medium.
In this `random walk', we just have to describe all the radiation sources, trace a path for each photon describing 
all interactions, and tabulate parameters of interest, such as intensity, flux, angle of exit, position of exit (for imaging), and wavelength.
These should converge to a mean and become statistically significant when a large number of photons are processed.
Many problems require iteration, and clever methods have been developed to handle this as well as high optical depths 
efficiently, as will be described later.   
In this section, we will describe the basic methods needed to solve a simple scattering problem, that of isotropic scattering 
in a plane-parallel atmosphere (see also Watson \& Henney 2001, and Gordon et al. 2001 for an overview
 of the MCRT scattering solution).
This is a problem that Chandrasekhar (1946, 1960) calculated analytically.  
His simplest case was a semi-infinite atmosphere, that is infinite in the $x, y,$ and $-z$ directions and photons emerge 
from the top of the atmosphere, defined at $z=0$.  This is our most time-consuming case, which can be approximated by a 
plane parallel atmosphere with a large optical depth ($\tau=7$ is sufficient) from bottom to top.   Coulson, Dave \& Sekera (1960) 
calculated finite thickness atmospheres using Chandrasekhar's method.  Our code can be tested by comparing to Coulson et al.'s tables, 
recently updated by Natraj, Li \& Yung (2009).

\subsection{The Fundamental Principle:  sampling probability distributions}

The essence of the Monte Carlo Method is sampling from probability distribution functions (PDFs), and this is referred to
as the `Fundamental Principle'.  To sample a quantity $x_0$ from a PDF $P(x)$, we need to invert the cumulative probability 
distribution (CPD), $\psi(x_0)$, which is the integral of $P(x)$:
\begin{equation}
\psi(x_0) = \frac{\int_{a}^{x_0}P(x)dx}{\int_{a}^{b}P(x)dx}.
\label{eq:fundprinc}
\end{equation}
As $x_0$ ranges from $a$ to $b$, $\psi(x_0)$ ranges from 0 to 1 uniformly (the proof of this can be found in 
Duderstadt \& Martin 1979; see also Kalos \& Whitlock 2008 or other standard Monte Carlo texts).  Thus, 
to sample a `random variate' $x_0$, we just need to call a random number generator that samples from 0 to 1 uniformly 
(we call this `uniform random deviate' $\xi$), and invert equation 1 to get $x_0$.  

To illustrate, we give the example of sampling the optical depth that a photon travels before being absorbed or scattered.  
The probability that a photon travels an optical depth $\tau$ without interacting is
\begin{equation}
P(\tau)d\tau = e^{-\tau}d\tau.
\end{equation}
Applying the fundamental principle:
\begin{equation}
\psi(\tau) = \frac{\int_0^{\tau_0}e^{-\tau}d\tau}{\int_0^{\infty}e^{-\tau}d\tau}=1-e^{-\tau_0} = \xi.
\end{equation}
Inverting this gives
\begin{equation}
\tau_0 = -\log(1-\xi),
\end{equation}
where $\xi$ is the uniform random deviate returned from the random number generator subroutine.   It is worth investigating the algorithm 
used by your compiler to find out how many numbers it generates before repeating.  A good source for a discussion on 
random number generators and a recommended algorithm is given in Numerical Recipes (Press et al. 2007).  

Sampling a scattering angle from an isotropic distribution ($P(\mu,\phi)d\mu d\phi =  d\mu/2 d\phi/ (2\pi)$) is also very straightforward,
giving
\begin{equation}
\begin{array}{l}
\mu_0 = 2 \xi_1 - 1 \\
\phi _0 = 2 \pi \xi_2
\end{array}
\end{equation}
where $\mu = \cos \theta$, $d\mu = \sin\theta d\theta$.

We discuss in Section 3 different methods for sampling from more complicated PDFs.  Kalos \& Whitlock (2008)
describe in detail different sampling methods.  Carter \& Cashwell (1975) describe methods
relevant to radiative transfer, such as sampling from a Planck function.

\subsection{The random walk}

To calculate this problem, we emit photons from the bottom of a plane-parallel atmosphere, defining $\tau_z=0$, and the top of the atmosphere is $\tau_z=\tau_{atm}$.  The initial photon position is $x,y,z = 0,0,0$, and the initial direction is $\mu_0, \phi_0=0$.   Sample optical depth from Eqn. 4, and move the photon to a new position:  $\tau_{znew} = \tau_{zold} + \mu * \tau$.  Check to see if $\tau_{znew}$ is greater than $\tau_{atm}$.  If not, sample direction from Eqn. 5 and continue to randomly walk until the photon exits.  When the photon exits the top of the atmosphere, tabulate its angle of exit.  Bin the angles uniformly between $\mu = 0-1$ and $\phi = 0-2\pi$:
\begin{equation}
i= integer (\mu N_\mu) +1
\end{equation}
\begin{equation}
j={integer}(\phi *N_\phi + 0.5) +1  ;
if j > N_\phi, j=1
\end{equation}
where ${integer}$ is a function that converts a real number to an integer (its actual call name depends on the computer language), and $N_\mu$ is the number of $\mu$ bins.  and $N_\phi$ is the number of $\phi$ bins.

\subsection{Calculating intensity and flux}

Next we want to calculate the intensity of the exiting binned photons.  From Chandrasekhar (1960; equation 1)
\begin{equation}
I_\nu = \frac{dE_\nu}{\cos \theta d\nu d\sigma dA dt}
\end{equation}
where $E_\nu$ is the energy at frequency $\nu$ exiting at an angle $\theta$ to the normal of a surface with area $dA$ into a solid angle $d\omega$ over time dt.   This describes a pencil beam of radiation emitted from the surface of the atmosphere.

If $N_{i,j}$ is the number of photons exiting at $\mu_i, \phi_j$,
and assuming for now monochromatic photons with no time dependence,
then the intensity $I_{i,j}$ is given by
\begin{equation}
I_{i,j} = \frac{h \nu N_{i,j}}{\mu_i \Delta\mu \Delta\phi dA}
\end{equation}
The intensity is usually normalized to flux $F$.
As defined in Chandrasekhar (1960), the net rate of flow of energy across a surface per unit area per unit frequency
interval is given by
\begin{equation}
\pi F = \int_{-1}^1 \int_0^{2\pi} I(\mu,\phi) \mu d\mu d\phi
\end{equation}
A total of $N_0$ photons are incident at cosine angle $\mu_0$, giving
\begin{equation}
\pi F = \frac{h \nu N_0}{\mu_0 dA}
\end{equation}
and therefore
\begin{equation}
\frac{I_{i,j}}{F} = \frac{\pi \mu_0 N_{i,j}}{\mu_i N_0 \Delta \mu \Delta \phi}
\end{equation}
If the incident radiation is isotropic, $I_\nu(\mu,\phi)=I_0$, then Eqn. 10 gives $F = I_0$.
According to Eqn. 8, $dE = I_0 \mu d\mu d\phi dA$.  Integrating over solid angle and area gives
$E = h \nu N_0 = \pi I_0$, which equals $\pi F$.  Substituting this definition of $F$ into Eqn. 9 gives

\begin{equation}
\frac{I_{i,j}}{F} = \frac{\pi N_{i,j}}{\mu_i N_0 \Delta \mu \Delta \phi},
\end{equation}
which is the same as that for parallel incident radiation except there is no factor of $\mu_0$.

By extending this algorithm to include electron scattering (Section 3.1), polarization, and albedo (Section 3.2.2), the code can
be compared to Chandrasekhar (1946, 1960) and Code (1950) for large optical depths, and Coulson et al. (1960) and 
Natraj et al. (2009) for varying optical depths and incident angles.  This is a great way to test out your Monte 
Carlo code, and learn how to compute intensity and flux.  When considering more complicated
problems with different boundary conditions, or frequency and time dependence, refer to the original definitions of 
intensity and flux to properly
normalize the results.  This is one reason I have
referred to Chandrasekhar's (1960) book many times over the last 30 years.  

\subsection{More complicated geometries}

The Monte Carlo Method solves problems in 3-D geometries as easily as 1-D, complicated scattering
functions as easily as isotropic, and low optical
depth more easily than high; therefore this is where it excels and is very complementary
to other methods.  All that is needed to solve any scattering problem is to describe where the photons originate from
and in what direction, where the scattering material is, how it scatters, and when the photon exits.  As described before, at
each scatter, a new photon direction is chosen and a new optical depth.  In most problems, the density of material varies
with position, and the distance a photon travels is related to the optical depth through the exinction opacity (the 
sum of the absorptive and scattering opacities) of the material:
\begin{equation}
d\tau = \chi_1 \rho ds = \chi_2 n ds = \chi_3 ds
\end{equation}
reflecting the different units the opacity might have.  In this case, the units of $\chi_1$ are cm$^{-2}$ g, 
the units of $\chi_2$ are cm$^2$ and the units of $\chi_3$ are cm${-1}$.  In the first case multiply by the
density $\rho$ (g cm$^{-3}$), in the second case by the number density $n$ (cm$^{-3}$), and in the third case,
the density has already been factored into the value of $\chi_3$.
As the photon propagates, equation 14 must be integrated either analytically or numerically.  The new photon
position is then calculated from
\begin{equation}
\begin{array}{l}
x = x_{old} + s \sin\theta \cos \phi  \\
y = y_{old} + s \sin \theta \sin\phi  \\
z = z_{old} + s \cos \theta
\end{array}
\end{equation}
In most problems where the density varies with position, we use grids to describe the problem, either
spherical-polar, cylindrical, or cartesian.  In each grid cell the density is constant across the cell.  Given
the photon propagation direction, the distance to the nearest wall is calculated, $s_{wall}$ (in a cartesian
grid, we find the distance to planes; in a spherical-polar grid, we find the distance to planes ($\phi$), cones ($\theta$) 
and spheres ($r$)).  The photon position is updated
using equation 15.  The optical depth is updated:

\begin{equation}
\tau = \tau_{old} + \chi \rho_{cell} s_{wall}
\end{equation}

If $\tau$ exceeds the sampled value (equation 4), the photon is moved back to where $\tau = \tau_0$; otherwise it
continues through the next cell where $x,y,z,$ and $\tau$ are updated again.  When $\tau = \tau_0$, the photon
scatters.

\subsection{Producing images}

Images are easily computed by tracking the position of the previous interaction.  When the photon exits,
its position of last interaction (scatter or emission)  is projected onto the $x-y$ plane perpendicular to the outgoing direction:
\begin{equation}
\begin{array}{l}
x_{image} = z_{old} \sin \theta - y_{old} \cos \theta \sin \phi - x_{old} \cos \theta \cos \phi  \\
y_{image} = y_{old} \cos \phi - x_{old} \sin \phi,
\end{array}
\end{equation}
where $( x_{old}, y_{old}, z_{old})$ are the coordinates of the last interaction.
Next we bin the photon into a pixel $(ix,iy)$ on the image:
\begin{equation}
\begin{array}{l}
ix = integer(nx (x_{image} + x_{max})/(2 x_{max})) + 1 \\
iy = integer(ny (y_{image} + y_{max})/(2 y_{max})) + 1,
\end{array}
\end{equation}
where $(nx,ny)$ are the number of $x$ and $y$ pixels in the image, and the image size ranges from $[-x_{max} : x_{max}]$ and  $[-y_{max}:y_{max}]$.

\subsection{Estimating errors}

In the simple case of isotropic scattering as described above, the photon energy
remains constant as it propagates through the medium, and the fractional error in the intensity is the 
Poisson statistical error  $1/\sqrt{N}$ where $N$ is the number of photons.
In more complicated problems as described below, if we sample properly the PDFs for scattering
and propagation, then the energy of each photon remains constant and is also given by simple Poisson statistical error.
As described below, we could sample from isotropic scattering and then weight the photon by its more complicated
phase function for scattering.  Then the errors can be estimated from the standard deviation of the summed intensities
of the outgoing photons normalized to $\sqrt{N}$.   When polarization is included, the other Stokes parameters
are estimated in the same way, by the standard deviation of the outgoing Stokes component (Q, U, or V), normalized 
to $\sqrt{N}$ (Wood et al. 1996).  The errors are minimized when the PDFs are sampled exactly. 
Gordon et al. (2001) also discuss error estimation.

\section{More complicated scattering problems}

The kinds of scattering problems usually investigated in astrophysics applications are electron, Compton, 
resonance line, and dust scattering.  In many cases, the scattering phase function (the angular dependence of 
the scattering function) can be
defined or approximated with analytic functions, and in other cases, they are computed numerically and 
described in tabular form.  All of these cases, including the polarization components, can be solved with relative 
ease with the Monte Carlo method.
I summarize one general method here, including polarization (see also Chandrasekhar 1960; Code \& Whitney 1995),
noting that there are other variations to implement this (Hatcher Tynes et al. 2001; Cornet, C-Labonnote, \& Szczap 2010; Hillier 1991).
We use the Stokes Vector $\bf{S}$ to describe the polarization:
\begin{equation}
{\bf S}(\theta,\phi) = [I(\theta,\phi),Q(\theta,\phi), U(\theta,\phi), V(\theta,\phi)]
\end{equation}
where $I$ is the intensity, $Q$ the linear polarization aligned parallel or perpendicular to the $z$-axis, $U$ is the linear polarization aligned $\pm45$\arcdeg to the $z$-axis and $V$ is the circular polarization.  The Stokes vector could also be defined as $[I_\parallel(\theta,\phi),I_\perp(\theta,\phi), U(\theta,\phi), V(\theta,\phi)]$, where $I_\parallel$ is the intensity of light with polarization parallel to the $z$-axis, $I_\perp$ has polarization perpendicular to the $z$-axis, and $Q=I_\parallel-I_\perp$.
A scattering diagram is shown in Figure 1 (Chandrasekhar 1960).
The photon is originally propagating into
direction ${\bf P_1}$ and will scatter into direction ${\bf P_2}$.  
In many scattering problems, the phase function can be described analytically dependent only on the angle $\Theta$
with respect to  ${\bf P_1}$.
For polarization problems, it is more complicated, because 
the polarization depends on the frame of reference.  We define the polarization in the ``observer's''  frame (the $x-y-z$
frame in Figure 1).
Thus, we need to rotate into and out of the photon propagation direction to apply the scattering matrix, using
Mueller matrices (Chandrasekhar 1960; Code \& Whitney 1995).
This is not strictly necessary, as the full scattering matrix can be calculated in the observer's frame (e.g., Whitney 1991a).
In magnetic problems, it is easier to define the scattering phase function with respect to the magnetic field
direction, and rotate in and out of these frames (Whitney \& Wolff 2002).  The resulting Stokes vector after scattering is:
\begin{equation}
{\bf S} = {\bf L}(\pi-i_2){\bf R} {\bf L}(-i_1){\bf S'},
\end{equation}
where ${\bf S'}$ is the incident
Stokes vector and ${\bf L}$ is the Mueller matrix that rotates in and out of the photon frame, defined as
\begin{equation}
{\bf L}(\psi) = \left[ \begin{array}{llll}
1 & 0 & 0 & 0 \\
0 & \cos 2\psi  & \sin 2\psi  & 0 \\
0 & -sin 2\psi & cos 2 \psi & 0 \\
0 & 0 & 0 & 1 \\ 
\end{array} 
\right]
.
\end{equation}
The scattering matrix ${\bf R}(\Theta)$ is
\begin{equation}
{\bf R}(\Theta) =
a 
\left[ \begin{array}{llll}
P_{11} & P_{12} & P_{13} & P_{14} \\
P_{21} & P_{22} & P_{23} & P_{24} \\
P_{31} & P_{32} & P_{33} & P_{34} \\
P_{41} & P_{42} & P_{43} & P_{44} \\
\end{array} 
\right]
.
\end{equation}
where $\Theta$ is the scattering angle measured from the incident photon direction and $a$ is a normalization
factor.
Note that if we want to ignore polarization, we can ignore all of the elements except $P_{11}$.

\begin{figure}
\centerline{\includegraphics[width=10cm]{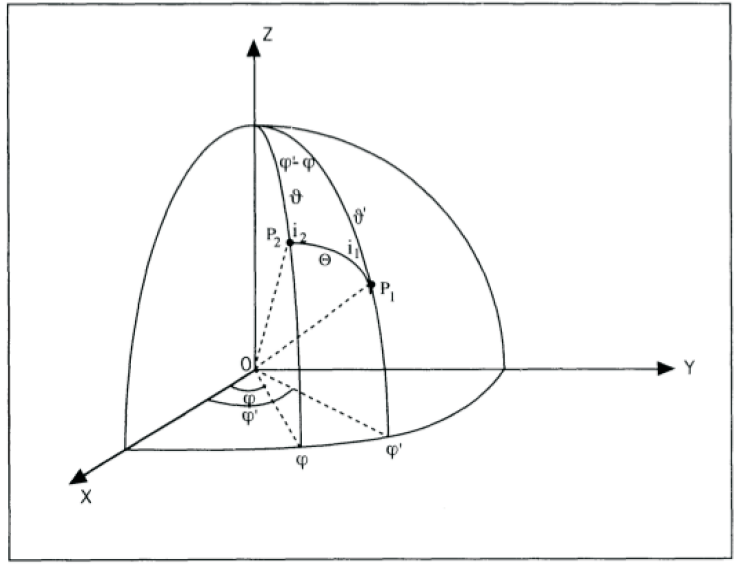}}
\caption{Geometry for scattering.  A photon propagating into direction $P_1$ ($\theta',\phi'$ in the observer's frame)
scatters through angle $\Theta$ into direction $P_2$ ($\theta, \phi$)}
\end{figure}

\subsection{Rayleigh scattering}

Let us consider the case of Rayleigh scattering, where 
\begin{equation}
\begin{array}{l}
a=3/4 \\
P_{11} = P_{22} = \cos^2 \Theta + 1 = M^2 + 1\\
P_{12} = P_{21} = \cos^2 \Theta - 1 = M^2 - 1 \\
P_{33} = P_{44} = 2 \cos \Theta  = 2 M \\
\end{array} 
\end{equation}
where $M = \cos\Theta$, and the other elements are 0.

Then the I Stokes parameter in the reference frame of the photon is computed:
\begin{equation}
{\bf S} = {\bf R} {\bf L}(-i_1){\bf S'},
\end{equation}
giving:
\begin{equation}
I =  ( M^2 + 1) I' + (M^2 - 1) \cos 2i_1 Q' - 2 M \sin 2 i_1 U'.
\end{equation}
We want to sample the scattering direction ($M,i_1$) from this function.

\subsubsection{Ignoring polarization and using lookup tables for sampling PDFs}
First consider the case where we ignore the polarization.  Then
 $I = I' (M^2 + 1)$.   There are a couple of ways we can sample
scattering angle from this PDF.
We could sample $M$ from a uniform angular
distribution (equation 5), and calculate a new photon intensity at each scatter from $I = I' (M^2 + 1)$.
Or we can sample the angle $M$ directly from the PDF $I = I' (M^2 + 1)$.  
In this case the photon intensity will always be equal to 1 as it propagates
through the medium.  
To do this for Rayleigh scattering, we apply the fundamental principle (equation 1),
\begin{equation}
\xi = \frac{\int_{-1}^{M_0} 1 + M^2 d M}{\int_{-1}^{1}1 + M^2 d M}= 1/2 + 3/8 M_0 + 1/8 M_0^2
\end{equation}
As described before, $\xi$ is a uniform random number between 0 and 1, obtained from a random number
generator.  Inverting equation 26 to get $M_0$ for each scatter is not trivial.  A fast way to sample $M_0$
is to make a table of the CPD (equation 26), $1/2 + 3/8 M_0 + 1/8 M_0^2$, which ranges from
0 to 1 uniformly.  Then linearly interpolate this table to get the value of $M_0$ that corresponds to the
value of $\xi$ obtained from the random number generator.
Once $M_0$ is computed, an azimuthal angle $i_1$ is sampled ($i_1 = 2 \pi \xi$), and the new direction
in the coordinate frame of the observer is computed (Fig. 1).

\subsubsection{Including polarization and using the rejection method for sampling from PDFs}

If solving the full polarization problem, we will sample $M$ and $i_1$ from the $I$ Stokes parameter calculated from equation 24.
As described before, we could sample $M (= \cos^2 \Theta)$ and $i_1$ from uniform angular
distribution (equation 5), and calculate a new photon intensity from equation 25.  Then the intensity of the photon
will vary as the photon propagates through the medium. 
For Rayleigh scattering, where the intensity varies only by a factor of 2 with
angle of scatter, it is okay to sample isotropically and weight the photon intensity.
For scattering that has a more peaked function, such as dust scattering, or in strong magnetic fields,
 this will lead to higher errors and systematic biases (many photons
with small intensities and few with large intensities but poor statistics).   To prevent this, I generally try
to sample from the exact probability distribution.  
A simple method that samples from complicated probability distributions is the rejection method.
All that is needed for this method is to know the peak of the PDF.

In the rejection method, we sample from a rectangle that encloses the curve of $P(x)$ vs $x$.  That is,
following equation 1, we sample $x$ uniformly from $a$ to $b$:  $x_0 = a + \xi (b-a) $; and we 
sample $y$ uniformly from $0$ to $P_{max}$, the maximum value of $P(x)$:  $y_0= \xi P_{max}$.  We ask if $y_0$ is less than $P(x_0)$.   If so, we accept $x_0$.  If not, we sample again.  It is like throwing random darts at the plot and only accepting those that fall below the curve.  By throwing enough darts, we accurately sample the different values of $x$ appropriately.  That is, in regions of the plot where $P(x)$ is low, we sample those values of $x$ less frequently than regions where $P(x)$ is large.
The rejection method is less efficient for highly peaked PDFs; that is, if the rectangle enclosing the PDF has a lot
of area above the PDF.  However, it is so simple to use that it is still usually much faster and easier to implement
than more complicated inversions of the CPD (equation 1).  
See Kalos \& Whitlock (2008) or other Monte Carlo texts for more examples and more sophisticated modifications to this method (such as enveloping highly-peaked functions with simple analytic highly-peaked functions which are sampled from first).

Going back to our scattering problem, as described in Figure 1, we want to sample scattering angles that change our direction from ${\bf P_1}$ to ${\bf P_2}$.   That is, we want to sample $\Theta$ and $\i_1$, compute the new Stokes
parameters and then rotate back into the observer's frame of reference.
Using the rejection method, we sample $i_1$ and $M= \cos \Theta$ from an isotropic distribution (equation 5):   $i_1 = 2 \pi \xi_1; M = 2
\xi_2 - 1$.   We calculate $I(M,i_1)$ from equation 25.  We sample $P(M,i_1) = \xi P_{max}$.   If $P(M,i_1)$ is
greater than $I(M,i_1)$, we accept $M$ and $i_1$ as our new scattering angles.  Otherwise, we resample until $P(M,i_1)$ is
greater than $I(M,i_1)$.  As mentioned previously, we need to know the value of $P_{max}$.  This can be determined analytically
or numerically (from brute-force calculation over all angles).  It is a good idea to verify that $P(M,i_1)$ never exceeds $P_{max}$ during the run.

Now that we have our new scattering angles $M$ and $i_1$, we compute the new propagation direction
and Stokes vectors in the observer's
frame.
The angles $i_2, \theta, \phi-\phi'$ (Figure 1) can be calculated from the spherical laws of sines and cosines (Green 1985).
The matrices are multiplied through and the Stokes parameters are calculated from equation 20.
The Stokes vectors are then normalized to the PDF we sampled from $P(M,i_1)$ (equation 25).
Then the $I$-Stokes parameter of the photon is equal to 1 as it propagates through.

\subsection{Dust scattering}

Since dust is ubiquitous throughout the universe, having the capabilities to solve the radiative transfer of dust
in multi-dimensional geometries allows us to model everything from planets, extrasolar planets, forming stars,
evolved stars, star forming regions, and galaxies throughout the universe.
Dust scattering can be approximated with analytic functions or tables produced from numerical models.

\subsubsection{Analytic functions}

The most famous analytic function is the Henyey-Greenstein (H-G) function (Henyey \& Greenstein 1941).  White (1979) added
to this with approximations for the polarization functions.  The elements of the scattering matrix  ${\bf R}(M)$ (where
$M = \cos \Theta$) (equation 22) are 
\begin{equation}
\begin{array}{l}
a=3/4 \\
P_{11} = P_{22} = (1 - g^2)/ (1 + g^2 - 2 g M)^{3/2}  \\
P_{12} = P_{21} =  - p_1 P_{11} (1-M^2)/(1+M2) \\
P_{33} = P_{44} = P_{11} (2 M)/(1+M^2) \\
P_{34} = P_{43} = -p_c P_{11} (1 - M_f^2)/(1+ M_f^2),
\end{array} 
\end{equation}
where $g$ is the scattering asymmetry parameter, ranging from 0 for isotropic scattering to 1 for fully forward scattering; $p_l$ is the maximum linear polarization; $p_c$ is the peak circular polarization; $M_f = \cos \Theta_f$, 
$\Theta_f = \Theta ( 1 + 3.13 s exp(-7 \Theta/\pi)$, and $s$ is the skew factor which we take to be 1 following
White (1979).   The other elements in equation 22 are 0.  Note that this function includes a circular polarization
component ($P_{34}$ and $P_{43}$).   This is a second order effect that depends on the linear polarization and is usually small.  

Multiplying through equation 24 to get the I Stokes parameter in the photon reference frame gives:
\begin{equation}
I = P_{11} I' + P_{12} \cos 2i_1 Q' - P_{12} \sin 2 i_1 U'
\end{equation}
The scattering angles $M,i_1$ can be sampled using the rejection method (Section 3.1.2).

If you don't care to solve the polarization problem, you just use $P_{11}$ for the scattering phase function.
This can be sampled from directly using the following formula (Witt 1977a):
\begin{equation}
M = \frac{1 + g^2 - [(1 - g^2)/(1 - g + 2g \xi)]^2}{2g}
\end{equation}
Witt (1977a) describes in detail a Monte Carlo dust scattering algorithm using this function, as well as superpositions
of H-G functions. He also describes how to force the first scattering in an optically thin nebula to make the code
more efficient (see also Gordon et al. 2001).

The other parameters that describe the dust properties are the extinction opacity $\chi$ (see equation 14) and the 
albedo $\omega$ (Section 3.2.2).
These as well as $g$ have been estimated observationally.  Theoretical models also match these as well as estimating
$p_l$ and $p_c$ which can be tested by comparing scattered light models to polarization observations.   All of these 
quantitites are wavelength-dependent.

\subsubsection{Dust scattering albedo}

The scattering albedo is the ratio of scattered to extincted (scattered + absorbed) flux, and it ranges from 0 to 1.  
This can be taken into account in one of two
ways:  either by weighting the photon at each scatter by the albedo, or by casting for a random number $\xi$ to determine if the photon is
absorbed or scattered at each interaction.   In calculations where we only consider the scattered component
of the radiation at a specified wavelength (e.g., dusty sources illuminated by UV, optical, and near-IR radiation),
we might think the first solution would be more efficient, that of weighting the scattered photons by albedo. 
This is often not the case, especially in sources with very high optical depths in some regions, where too
much computing time is wasted on photons with little weight and therefore little contribution to the final answer.
In those cases, it is much faster to let the photon scatter or absorb by casting for a random number.
If $\xi$ is less than the albedo, the photon scatters; otherwise, it is killed, and we proceed to the next photon.

\subsubsection{Tabular functions}

The scattering matrix ${\bf R}(M)$ (equation 22) can also be computed numerically.  Tables of the 16-element
matrix as a function of scattering angle are read in at the beginning of the computation.  For spherical grains,
the matrix is simplified, with only 4 independent elements needed, as above in the analytic approximation.
For randomly oriented non-spherical grains, 6 independent elements are needed.  For aligned grains, all
16 elements are non-zero.

The rejection method works well at sampling the tabular functions.  At the beginning of the code, the peak
of the $M_{11}$ element is computed, which we will call $I_{peak}$.  In the cases I have tried, this is
also the peak of the $I$ Stokes vector even when the incident radiation is polarized.  At each scatter, 
as described in \S 3.1.2, the angles $M$ and $i_1$ are sampled uniformly.  The values of $P_{11}$, $P_{12}$, and $P_{13}$, $P_{14}$ (if
non-zero) are calculated by interpolating the tables (which depend on $M$).  Then the I Stokes parameter \ in the reference frame of the
photon is computed from equation 24:
\begin{equation}
I = P_{11} I' + (P_{12} \cos 2i_1 + P_{13} \sin 2i_1) Q' + (P_{13} \cos 2 i_1 - P_{12} \sin 2 i_1) U' + P_{14} V'.
\end{equation}
For spherical grains, only 8 of the scattering matrix elements are filled with 4 unique elements, as in the analytic
prescription above:  $P_{11} = P_{22}$, $P_{12} = P_{21}$, $P_{33} = P_{44}$, $P_{34} = P_{43}$, and the rest
are zero, giving the same form as equation 28.
 As described in Section 3.1.2, a random number $\xi$ is chosen between 0 and the peak of I; if $\xi$ is less 
than $I_{peak}$, the angles $M$ and $i_1$ are accepted, and the rest of the Stokes vectors are calculated from equation 20.  
To verify that we properly calculated the peak of the scattering function,
we check at each scattering that the I Stokes parameter does not exceed $I_{peak}$.  if it does, we need to rerun the code with the correct value.
Once the scattering angle has been calculated, the other angles are computed, and the Stokes vector
in the observer frame are computed (equation 20), as described in Section 3.1.2.

\subsubsection{Aligned grains}

Aligned grains use the full 16-element scattering matrix, calculated as described in the previous section (Section 3.2.3).  Instead
of rotating in and out of the photon direction frame, we rotate into and out of the frame aligned with the magnetic field along the $z$-axis.  The 16-element scattering matrix is defined with respect to field direction rather than photon direction.
The additional component here is in the random walk, where the opacities depend on the polarization of the photon.
Photons traversing the medium develop Q polarization in the frame of the magnetic field, called dichroism.  Photons with some U polarization (w.r.t. magnetic field direction) develop V polarization, called birefringence.
Whitney \& Wolff (2002) describe how to implement these effects along the photon propagation
path.

\subsubsection{Applications of continuum scattering problems}

Most electron scattering applications are in resonance line scattering of stellar winds, as described
in the next section.  Whitney (1991a) described how to calculate the scattering of electrons in
magnetic fields of arbitrary strength, and showed how the magnetic effects can explain the unusual
polarization behavior in the polarization of magnetic white dwarfs (Whitney 1991b).

The most widespread applications of Monte Carlo (MC) continuum scattering have been for dust scattering.
Witt (1977a,b,c) and Witt \& Oshel (1977) pioneered this field describing algorithms for sampling the Henyey-Greenstein
function and computing the MC radiative transfer.  Witt and collaborators applied these
codes to galaxies showing the ``blueing'' due to scattering partially compensates for reddening
by extinction (Witt, Thronson \& Capuano 1992) and the effects of clumping on the radiative
transfer (Witt \& Gordon 1996, 2000).
Bianchi et al. (2000) also studied the effect of clumping in dusty galaxies.
Boisse (1990) studied the effects of clumps in the penetration of UV photons inside molecular
clouds.  
Whitney \& Hartmann (1992, 1993), Kenyon et al. (1993), and Fischer, Henning \& Yorke (1994) calculated dust scattering and 
polarization in 2-D structures--disks, envelopes, and bipolar cavities---surrounding protostellar envelopes.  Several authors 
have modeled high spatial-resolution  images from Young Stellar Objects (YSOs),
determining disk/envelope properties and grain size distributions
(e.g., Wood \& Whitney 1998; Cotera et al. 2001; Schneider et al. 2003; Wolf, Padgett \& Stapelfeldt 2003; Watson \& Stapelfeldt 2004, 2007; Duchene et al. 2004; Stark et al. 2006; Watson et al. 2007 and references therein),
and polarization maps (Whitney, Kenyon \& Gomez 1997; Lucas \& Roche 1997, 1998).
Whitney \& Wolff (2002), Lucas (2003), and Lucas et al. (2004) modeled polarization maps of YSOs with aligned
grains, to study the magnetic field structures.
Jonsson (2006) describes a code for computing scattering in galaxies.  The advances of this
code are that it follows a spectrum of photons through, rather than a single wavelength;
and is designed to work with SPH simulations and on an adaptive grid.  This code is widely
used in the study of galaxy evolution to visualize galaxy images produced from SPH simulations
(such as the GADGET code; Springel, Di Matteo \& Hernquist 2005).

\subsection{Line scattering problems}

\subsubsection{Resonance line scattering and scattering in flows}

Resonance lines are transitions to and from the ground states of bound electrons.  The scattering matrix is the sum of a 
Rayleigh phase function plus an isotropic function.
In a flow, such as an expanding atmosphere or universe, we take into account the Doppler shifts of
the fluid with respect to the incident photons. 
Hillier (1991) calculated the electron scattering of lines in Wolf-Rayet stars.
He described how to calculate the emission location, that is, where the photon of a given direction
and frequency will resonantly interact with the flow, and how to transform the frequency from one
frame to the next in the flow.
Kurosawa \& Hillier (2001) applied these algorithms in a 3-D tree-structured grid, and demonstrated
their model on interacting winds in massive binaries (see also Kurosawa, Hillier \& Pittard 2002 for an application
to the massive binary V444 Cyg).
Sundqvist, Puls \& Feldmeier (2010) calculate resonance line formation in 2-D wind models, in an
ongoing effort to resolve a very interesting new controversy on mass-loss rates from clumpy
massive stellar winds (see Puls, Vink \& Najarro 2008).  They require higher mass loss rates than in the optically thin clump models which they say resolves the controversy.
Knigge, Woods \& Drew (1995) calculated resonance line scattering in accretion disk winds.

Another useful application for resonance line scattering is the radiative transfer of Ly$\alpha$
photons.  This problem can be approximately solved analytically only for a limited number of
cases such as a static, extremely opaque and plane-parallel medium.  Several authors describe
radiative transfer calculations (e.g.,  Zheng \& Miralda-Escude 2002; Verhamme, Schaerer \&
Maselli 2006; Laursen et al. 2009))
and apply them to, e.g., Ly$\alpha$ radiative transfer in a dusty, multiphase medium (Hansen \&
 Oh 2006), Ly$\alpha$ pressure in the neutral intergalactic medium (Dijkstra \& Loeb 2008),
 Ly$\alpha$ escape fractions from simulated high-redshift dusty galaxies (Laursen, Sommer-Larsen
 \& Andersen 2009), cosmological reionization simulations (Zheng et al. 2010), and the Ly$\alpha$ forest
 around high redshift quasars (Partl et al. 2010).

\subsubsection{Relativisitic scattering}

In principle, the calculations for relativistic scattering processes are similar, with additional transformations of the photon frequency
in and out of the co-moving frame.
If gravitational redshift is important, we need to apply this to the photon frequency at each step of the photon path integration.
For more information, I refer the reader to other authors who know much more than I:  Wang, Wasserman \& Salpeter (1988) calculate cyclotron line resonance transfer in neutron star atmospheres; Fernandez \& Thompson (2007) also calculate cyclotron resonance scattering in 3-D geometries. Stern et al. (1995) describe a large particle (LP) method for simulating non-linear high-energy processes
near compact objects.  And Dolence et al. (2009) describes a general code (grmonty) for relativistic radiative transport.

\section{Including emission}

Adding emission usually adds wavelength dependence to the problem and allows us to model the spectral dependence
of an astrophysical source.  The dominant emission processes are from gas and dust.  We start with dust, which
is the easiest to calculate.  Fortunately, a wide variety of astrophysical problems can be addressed with 3-D
dust radiative transfer, due to the wealth of infrared data recently available from, e.g., the Spitzer Space Telescope, Herschel Space Observatory, Wide Field Infrared Survey Explorer (WISE), and the upcoming James Webb Space Telescope.

\subsection{Dust radiative equilibrium}

Due to the nature of its opacity, dust generally scatters and absorbs optical radiation, and emits infrared
radiation.   For grains larger than about 200 A in radius, we can usually assume that the dust is in thermal
equilibrium with the surrounding gas (we will address
smaller grains in Section 4.2).  The gas-to-dust mass ratio is about 100 in our Galaxy.  Even though there is much more 
gas mass than dust, its opacity is
many orders of magnitude larger than gas, so we can usually neglect the gas opacity in dusty nebulae.

We calculate the radiative transfer as described previously, but when a photon is absorbed (see Section 3.2.2), we re-emit
a thermal photon.  To do that, we need to know the temperature of the dust.  This is straightforward to solve under conditions 
of radiative equilibrium and local thermal equilibrium (LTE).  
The radiative equilibrium process describes the condition when all of the energy is transported by radiation.
Then we can say that the total energy 
absorbed by a given volume of material is equal to the total
energy emitted (Mihalas 1978):
\begin{equation}
4 \pi \int_0^\infty \chi_\nu (S_\nu - J_\nu) d\nu,
\end{equation}
where $S_\nu$ is the Source function, or the ratio of the total emissivity to the opacity, $J_\nu$ is
the average intensity in the same volume, and $\chi_\nu = \kappa_\nu + \sigma_\nu$ is the mass extinction
coefficient. 
In local thermal
equilibrium, we can write (Mihalas 1978):
\begin{equation}
S_\nu = (\kappa_{\nu} B_\nu + \sigma_{\nu} J_\nu)/(\kappa_{\nu} + \sigma_{\nu})
\end{equation}
where $\kappa_{\nu}$  and $\sigma_{\nu}$ are the mass absorption and scattering coefficients, respectively, and
their sum is $ \chi_\nu$ (in units of cm$^2$/g).  The condition of radiative equilibrium is then
\begin{equation}
\int_0^\infty \kappa_{\nu} B_\nu (T) d\nu = \int_0^\infty \kappa_{\nu} J_\nu d\nu,
\end{equation}
This is all the information we need for our Monte Carlo calculation.  We will do our calculation on a 
grid so we can calculate the volume and mass of each cell for the emission properties.  
This also allows flexibility in including arbitrary
density functions and makes optical depth integrations straightforward (Section 2.4).  

Bjorkman \& Wood (2001, hereafter BW01) describe how to determine the temperature of each grid cell by equating the total
absorbed photons with those emitted assuming thermal equilibrium.  This gives
\begin{equation}
\sigma T^4_{cell} = \frac{N_{cell} L}{4 N \kappa_P(T_{cell}) m_{cell}},
\end{equation}
where $N_{cell}$ is the number of photon packets absorbed in the cell, $L$ is the source luminosity,  $\kappa_P(T_{cell})$
is the Planck mean opacity, $m_{cell}$ is the mass of the cell, and $N$ is the total number of photon packets in the simulation.
This applies to any continuous opacity source that is independent of temperature.  To solve this equation efficiently,
we pretabulate the Planck mean opacities and use a simple iterative algorithm.

When a photon is absorbed in a cell we sum its energy into an array for use in computing equation 34.
We then emit a new photon of equal energy to conserve radiative equilibrium.  All that's required is to properly
sample its frequency from the emissivity function converted to a PDF:
\begin{equation}
\frac{dP_{cell}}{d\nu}  =  \frac{j_\nu}{\int_0^\infty j_\nu d_\nu}  = \frac{\kappa_\nu B_\nu(T_{cell})}
{\int_0^\infty \kappa_\nu B_\nu(T_{cell}) d\nu}
\end{equation}
where $(dP_{cell}/d\nu)$ is the probability of emitting a photon between frequencies $\nu$ and $nu + d\nu$.  We precompute the running
integral of this function (that is, the cumulative probability distribution or CPD, see Section 3.1.1) for a range of frequencies and temperatures,
and interpolate the table based on the
sampled random number $\xi$ to get $\nu$.

At the start of our simulation, we do not know the temperature of each cell, so we use an arbitrary value
(we start with 3 K), and use the absorbed photons to determine the temperature. 
We can iterate, i.e., do the calculation
several times, and calculate a new temperature for each cell (equation 34) after each iteration, until the cell
temperature converges (Lucy 1999a).
Alternatively, we can correct the temperature as we go and emit from a corrected emissivity spectrum (BW01). This corrects the emitted spectrum so that the total emitted spectrum at the end of the simulation is appropriate
for the temperature of that cell.  For example, if the cell starts out cold, the emitted photon frequencies will be lower
than the proper spectrum, so as the temperature warms up, we will sample from an overly ``hot" spectrum to emit
higher frequency photons.   This is described graphically in Figure 1 of BW01.
The temperature correction probability distribution is
\begin{equation}
\frac{dP_{cell}}{d\nu}  = \frac{\kappa_\nu}{K} \left (  \frac{dB_\nu}{dT}  \right )_{T=T_{cell}},
\end{equation}
where $K = \int_0^\infty \kappa_\nu (dB_\nu/dT) d\nu$ is the normalization constant.  Again, we can precompute the CPD and interpolate
from this to sample $\nu$ based on random number $\xi$.

Lucy (1999a) derived a much faster way to compute the total absorbed radiation in a grid cell (the right-hand-side of equation 33), 
using the pathlengths
of {\it all} photons crossing a cell, rather than summing only those absorbed.   This gives
\begin{equation}
 \int_0^\infty \kappa_{\nu} J_\nu d\nu = \frac{L}{4 \pi N V} \sum \kappa_\nu l,
\end{equation}
where $V$ is the volume of the cell, $l$ is the pathlength across the cell that a given photon traveled, and the others
are as defined in equation 34.
The pathlengths are summed during the optical depth integration as the photon travels through various
cells on its way to an interaction.
Following BW01 and equating this with the emitted radiation to solve for temperature,
we get:
\begin{equation}
\sigma T^4_{cell} = \frac{ \rho_{cell} L \sum \kappa_\nu l }{4 N \kappa_P(T_{cell}) m_{cell}},
\end{equation}
Thus, we can call our temperature solver with $\rho_{cell} \sum \kappa_\nu l$ in place $N_{cell}$.
Robitaille (2011) presents a variation of this method where he tabulates the specific internal energy rather
than the temperature. This allows the straightforward inclusion of other heating sources in addition to LTE dust.  
The temperature can be calculated if needed from the internal energy, similar to
equation (38). 
The simplest way to implement the Lucy method for correcting the temperature is with an iterative scheme.
The temperature remains constant during an iteration, we sample frequency 
from the emissivity (equation 35), and then calculate a new temperature for each cell at the end
of the iteration (equation 38).
Lucy (1999a) notes that this temperature correction 
scheme appears identical to the ``notorious" lambda-iteration procedures that are known to fail (Mihalas 1978);  however it is not the same, because flux is conserved exactly across all
surfaces.  In fact, this method converges in only a few iterations (3-4).

This method has several advantages over BW01:  1)  It is very fast at converging the temperature.   Chakrabarti \& Whitney (2009) quantified this by running several 3-D simulations and comparing the BW01 and Lucy methods.
In the Lucy iteration method, the number of photons required to get an accurate temperature is approximately $N_{temp} \sim 2 N_{grid}$, where $N_{grid}$ is the number of grid cells.  The BW01 method requires at least $N_{temp} \sim 100 N_{grid}$.  In the Lucy method, we run the first $n$ iterations using $N_{temp}$ photons, and then run the final iteration using  $N_{SED}$, the number of photons required to produce an SED of our desired signal-to-noise.  Usually, $N_{SED}$ is much larger than $N_{Temp}$.  In 2-D problems, the run-time of Lucy and BW01 is similar;
in 3-D problems, because there are so many more grid cells, the Lucy method runs much faster.
Robitaille (2011) describes a robust method to determine convergence.
2)  The Lucy method is easily parallelizable.  Since the temperature remains constant during an iteration, the
photons can be divided up among several processors and run independently.  At the end of each iteration,
they are summed up and a new temperature is calculated.
Robitaille (2011) shows the speedup expected as a function of number of processors.
3) More complicated physical processes that require iteration can be incorporated in a straightforward way. For example,
including temperature dependent opacities (e.g., gas opacity); calculating grain alignment from moments of the radiation intensity; and
calculating non-thermal small grain emissivity which requires knowledge of the average intensity in a grid cell.

\subsubsection{High fidelity spectra and images}

A useful technique for computing a high signal-to-noise image and SED is to `peel-off' a photon in a specified
(observer's) direction at every interaction (Yusef-Zadeh, Morris, \& White 1984).  When a photon is initially emitted, in addition to its sampled
direction, we emit an additional photon into one or more specified observer directions, weighted by
the PDF, or the probability that it would have gone in this direction.  The photon's intensity is additionally
weighted by the extinction it undergoes on its way to the observer $I = I_0 e^{-\tau}$ where $\tau$ is the
integrated optical depth along its path.   At each interaction (scattering or emission), we
again peel-off a photon into the observer direction, weighted by the PDF (for scattering or emission), and
the extinction.  Note that the peeling-off technique does not replace the regular Monte Carlo simulation, but is
an added computation.  The main `trick' with this is that we have to make sure that the peeled photon is normalized
properly.  In the regular simulation, this is done at the end of the simulation with the conversion of exiting
photons to flux and energy; during the simulation, the PDF's are normalized to range from 0-1 (to match
the random number range).  For example, in emitting photon packets
from a limb darkened star, we emit each photon with the same energy, but the distribution of emitted photons
varies with angle.  For the peeled photon, we weight it by the limb-darkening law and need to normalize it
properly.  Fortunately, this is easy to verify by comparing the peeled images and spectra with the regular
Monte Carlo in simulations that test all the emission and scattering processes (i.e., viewing images and SEDs of the
star only, then scattering-only
simulations, emission-only, high and low optical depths, etc.).


\subsection{The diffusion method}

In sources with high optical depths, MCRT can become very slow to compute when the photon path length is much shorter than the escape length from a given region.   In dust radiative transfer this effect is offset to some extent because the opacity of dust decreases with increasing wavelength: optical photons that are absorbed and re-emitted by the cooler dust get converted to infrared photons that can usually escape. Thus sources with visual optical depths of even 1000 are computed quickly.  However, in regions of much higher optical depths, such as protostellar disks, the photons effectively get trapped in the disk midplane, undergoing millions of interactions before escaping.
Min et al. (2009, hereafter M09) developed a modified random-walk (MRW) that moves photons through optically thick regions, using the diffusion approximation.

In the MRW method, when the optical depth in a grid cell is much larger than 1, we define a sphere whose
radius is smaller than the distance to the closest wall, and travel to the edge of the sphere in a single step.
The true distance the photon would have traveled in a random walk is calculated using the diffusion approximation.
This along with the average mass absorption coefficient are used to compute the total energy deposited and therefore
the temperature of the cell.  A new photon emerges from the sphere with the frequency sampled from the Planck
function at the local dust temperature.  
If the BW01 temperature correction method is used, the photon frequency is sampled from $d B_\nu(T)/dT$.
Robitaille (2010) showed how to compute the local diffusion coefficient D, the average
mass absorption coefficient and the dust emission coefficient $\eta_\nu$ without iteration, giving:
\begin{equation}
D=\frac{1}{3 \rho \overline{\chi}_R},
\end{equation}
\begin{equation}
\overline{\kappa} = \frac{\int_0^\infty \kappa_{\nu} B_\nu (T) d\nu}{\int_0^\infty B_\nu (T) d\nu} = \overline{\kappa}_P,
\end{equation}
\begin{equation}
\eta_\nu = \chi_\nu B_\nu (T) \frac{\overline{\kappa}_P}{\overline{\chi}_P},
\end{equation}
where $\chi_P$ is the Planck mean opacity,
\begin{equation}
{\chi}_P =  \frac{\int_0^\infty \chi_{\nu} B_\nu (T) d\nu}{\int_0^\infty B_\nu (T) d\nu} 
\end{equation}
and ${\chi}_R$ is the Rosseland mean opacity:
\begin{equation}
\frac{1}{\chi_R} =  \frac{\int_0^\infty \chi_{\nu} B_\nu (T)/\chi_{\nu}  d\nu}{\int_0^\infty B_\nu (T) d\nu}.
\end{equation}
Robitaille (2010) describes the implementation
of the MRW algorithm in his Section 3, so I refer the reader to that.

M09 also describe a Partial Diffusion Approximation (PDA) which can be used to obtain a reliable temperature
in regions where few if any photons reach, such as the midplane of an externally illuminated disk with
no self-luminosity due to accretion. 
For computations of images and SEDs, if no photons reach a given
region, none are emitted, so PDA is not needed.  However, if we want to solve for the vertical hydrostatic
density distribution of the disk, the temperature in all regions is required.
The PDA assumes  that no photons escape the optically thick region without interactions, which simplifies
the 3-D radiative diffusion equation (Wehrse, Baschek  \& von Waldenfels 2000; Rosseland 1924)
\begin{equation}
\nabla \cdot (D \nabla E) = \frac{1}{c} \frac {\partial E}{\partial t}
\end{equation}
to
\begin{equation}
\nabla \cdot (D \nabla T^4)  = 0. 
\end{equation}
This results in a system of linear equations that can be solved knowing the temperature at the boundaries
of the optically thick regions (Robitaille 2011 shows how this can be solved on a spherical polar grid).
Thus the PDA requires iteration, using the temperature calculated from
the MCRT solution.  The PDA overestimates the temperature slightly because it does not take into account
the few very long-wave photons that can escape from the region and cool it more efficiently.

\subsection{Non-equilibrium dust (small grain emission)}

Grains smaller than about 200 A, or Very Small Grains (VSGs), as well as large molecules such as Polycyclic Aromatic Hydrocarbons (PAHs) undergo quantum heating from even single photons, which leads to temperature fluctuations.  
These fluctuations depend on the size of the particle. Given a probability distribution P(T)dT for the temperature of a grain,
the emission from an ensemble of VSGs is given by (Misselt et al. 2001)  
\begin{equation}
L(\nu) = 4 \pi \sum_i \int_{a_{min}}^{a_{max}}  n_i(a) \sigma_i(a,\lambda) da \int B_\nu (T_{i,a}) P(T_{i,a}) dT
\end{equation}
where $i$ is the species of the grains (e.g., silicates or carbon), $n$ is the number density of grains (typically units are
cm$^{-3}$) of radius $a$
and $\sigma$ is the cross section of the grains (in units of cm$^2$).  
This can be compared to the left-hand side of equation 33, where the grain cross sections are
already integrated over size and are all assumed to emit at the same temperature T, which is valid for
large grains.
Misselt et al. (2001) describe how to determine P(T) for VSGs using the continous cooling approximation developed
by Guhathakurta \& Draine (1989), which speeds up the calculation significantly.
They describe an even more simplied approach to compute the PAH emission:
\begin{equation}
L_{PAH}(a,\nu) = 4 \pi \sigma(a,\nu) \overline{B[T(t)]},
\end{equation}
where the Planck function is averaged over the mean time between absorptions calculated from 
\begin{equation}
\frac{1}{\overline{t}} = \frac{4 \pi}{h c} \int_0^{\nu_c} \sigma(a,\nu) J_\nu d\nu,
\end{equation}
where $\nu_c$ is the cutoff frequency in the optical/UV cross section of the PAH molecule (Desert et al. 1990).

In their radiative transfer algorithm, Misselt et al. (2001) first process the stellar
and nebular sources, calculating the transmitted, scattered and absorbed photons in the grid.
Then they calculate the dust emission and transfer based on the heating from the absorbed
photons.  They iterate on the fractional change of energy absorbed by the grid.  This method does not
conserve energy in a given iteration and may be subject to Lambda iteration issues.   
The large grain emission is as described in the radiative equilibrium equation 33, using
the average intensity of each cell computed at the end of an iteration.  The PAH and very small grain component is as 
given in equations 46 and 47.  The solution for the very small grains is the most computationally expensive
part of the code.

Pontoppidan et al. (2007) also use the method of  Guhathakurta \& Draine (1989) to compute the heating of the very small
grains, and do not compute the PAH emission (though they do include PAH absorption opacity).  Photons absorbed by these very small grains are lost in the first iteration, to be released in a post-processing step and/or in a second iteration.

Wood et al. (2008) bypass the temperature calculations of the VSGs and PAHs altogether, and use look-up tables
for the emissivity of these species.  The input to the lookup tables is the average intensity $J$ in each
grid cell, calculated using the Lucy (1999a) method (equation 33, without the opacity). This method requires iteration.  In each iteration the photons are emitted from the star and other luminosity
sources (e.g., disk accretion) and are processed as described in previous sections.
At each interaction, we sample a probability that a photon is absorbed by a thermal grain, a VSG, or a PAH molecule,
based on the relative opacities of these material for the frequency of the incoming photon.
If a thermal grain, a thermal photon is emitted based on the temperature of the cell (equation 35); if a VSG or PAH, a non-thermal photon is emitted from the pre-computed emissivity spectra based on $J$ in the cell.
After each iteration a new temperature and $J$ are computed in each cell.  
Energy is conserved, and the models converge in 3-4 iterations.
This method is as fast as the radiative equilibrium method using the Lucy method.
The lookup tables incorporate
all the physics of the temperature fluctuations and emission as a function of input radiation field, but are
pre-computed so that it does not slow down the radiative transfer calculations.
The main approximation to the Wood et al. (2008) implementation is that they 
do not take into account the frequency dependence of the average intensity ($J_\nu$).  This assumption is not as egregious as it might
seem because the wavelength dependence of the opacity is taken into account, ensuring that PAH and VSG
photons are not emitted in regions with high $J$ but low probability of excitation.  
Robitaille (2011) improves on this method by tabulating the energy absorbed by the PAH and VSGs rather
than the average intensity.  This better samples the spectral shape of the emissivity; that is, if the intensity
peaks in the UV, the energy absorbed would predict a higher excitation emissivity.

\subsection{Aligned grain emission}

Thermal emission from aligned grains is similar to that of spherical grains except the full Stokes matrix is used in the emission.  The dust opacities need to be calculated, along with the degree of alignment.   Fiege \& Pudritz (2000) describe a method for
emitting polarized submillimeter emission in molecular clouds.  
 Bethell et al. (2007) and Pelkonen, Juvela \& Padoan (2009) show how to calculate the degree of alignment using radiative torques.  
Hoang \& Lazarian (2008, 2009a, 2009b), and Hoang, Draine \& Lazarian (2010) present new calculations on the radiative torque mechanism.  
 Because of the low opacities at these wavelengths, the absorption and
scattering is ignored in these calculations.  
In protostellar disks where the grains are larger and the optical depths higher, these approximations
are likely not valid.  Whitney \& Wolff (2002) describe how to include absorption along the photon path and scattering
of aligned grains.   When emission, scattering, and absorption are included, models can be made at all wavelengths
and densities.

\subsection{Applications of dust MCRT}

Several authors have developed dust MCRT codes that can be applied to a variety of astrophysical objects.
Their methods are generally similar to what I described above but there are variations in, for example, 
conserving energy by re-emitting photons as they are absorbed vs separating the initial emission and re-emission
processes; or different coordinate-system rotations for the Stokes vectors (conceptually simple vs computationally
efficient).  Numerical techniques and codes have been described by Lucy (1999a), 
Wolf, Henning \& Stecklum (1999), Wolf \& Henning (2000), Misselt et al. (2001), Bjorkman \& Wood (2001), Wolf (2003),  Stamatellos \& Whitworth (2003), Stamatellos, Whitworth, \& Ward-Thomson (2004), Whitney et al. (2003a,b), Niccolini
et al. (2003), Goncalves, Galli \& Walmsley (2004), Baes et al. (2005), Pinte et al. (2006), Niccolini \& Alcolea (2006), Pontoppidan et al. (2007), Bianchi (2008), Wood et al. (2008), Min et al. (2009), Kama et al. (2009),
and Robitaille (2010, 2011).
Adaptive grid techniques have been described by Niccolini \& Alcolea (2006).
Benchmark tests have been made by Pascucci et al. (2004) and Pinte et al. (2009).

These codes have been applied widely in the study of protostellar envelopes/disks, and galaxies.  In both
cases, clumpy structures (e.g., Schartmann et al. 2008 and Bianchi 2008 for galaxies,  Indebetouw et al. 2006 for
protostars, Doty, Metzler, \& Palotti 2005 for externally heated molecular clouds), and other asymmetric dust distributions (e.g., outflow cavities and disks) require 2-D and 3-D radiative transfer
codes to properly interpret the SEDs, images, and polarization. 

Grain alignment models have been applied to near-IR polarization maps, to determine magnetic
field structures in protostars (Whitney \& Wolff 2002; Lucas 2003; Lucas et al. 2004); and to submillimeter polarization maps to determine
magnetic structures (Fiege \& Pudritz 2000), density distributions, grain size distribution (Pelkonen
et al. 2009), and to test the radiative torque theories for grain alignment, polarization-Intensity
relations (Bethell et al. 2007; Pelkonen, Juvela \& Padoan 2007), and the Chandrasekhar-Fermi formula (Padoan et al. 2001).

The recent explosion of optical and IR data from several observatories and surveys (e.g., Spitzer Space Telescope, Herschel
Space Telescope, Hubble Space Telescope, 2MASS, UKIDDS, WISE), combined with advances in dynamical simulations that 
provide realistic density distributions, has made the development of 3-D dust radiative transfer a very fruitful area of research.  

\subsection{Gas emission}


\subsubsection{Non-LTE MCRT and flows}

As in the scattering  and dust emission processes, MCRT is very complementary to other methods.
Whereas traditional methods excel in high optical depth LTE 1-D geometries, MCRT can excel in
non-LTE, 3-D geometries with complex velocity fields and anisotropic radiation
fields.  
Bernes (1979) outlined a procedure for non-LTE multi-level radiative transfer and demonstrated the method
for CO line profiles in a spherical, homogeneous, collapsing dark cloud.
Since then, several authors have improved on the Bernes (1979) algorithms to, e.g.,
extend to 3-D (Park \& Hong 1995) allow for
very high optical depths (Hartstein \& Liseau 1998), treat clumpy structures (Park, Hong \& Minh 1996; 
Juvela 1997;
Pagani 1998), accelerate the convergence and include dust emission Hogerheijde \& van der Tak (2000),
and include multiple molecules (Pavlyuchenkov et al. 2007).

The application of MCRT to the computation of expanding gaseous envelopes was described by
Abbott \& Lucy (1985).    Mazzali \& Lucy (1993) adapted this code to supernova envelopes, where a
single continuum photon can interact with many more spectral lines due to the
high velocities of the outflow ($\sim 30000$ km s$^{-1}$).
The Monte Carlo approach is better suited to this problem than the formal integral
type solutions.  Mazzali \& Lucy (1993) include ionization, electron scattering and line scattering
in their code. Lucy (1999b) improves the line formation treatment of this code and the
noise in the emergent spectrum by using the formal integral for the emergent intensity.
Lucy (2005) removes many of the simplifying assumptions in the earlier codes and
solves the time-dependent 3-D NLTE transfer in homologously expanding ejecta of a SN,
given the distribution of mass and composition  at an initial time $t_1$.
Kasen, Min \& Nugent (2006) describe a similarly capable code, which also includes polarization and non-grey 
opacities, that can provide direct comparison
between multidimensional hydrodynamic explosion models and observations.
Maeda, Mazzali \& Nomoto (2006) and Sim (2007) also developed similar codes based on the Lucy methods.

Long \& Knigge (2002) apply the methods of Mazzali \& Lucy (1993) to calculate line formation and transfer
in accretion disk winds.  Sim, Drew \& Long (2005) extended this code to include `macro atoms',
as devised by Lucy (2002, 2003), allowing energy conservation and radiative equilibrium to be enforced at all times.
This allows lines formed by non-resonance scattering or recombination to be modeled.

Carciofi \& Bjorkman (2006) employ a 3-D non-LTE code to study the temperature and ionization
structure of Keplerian disks around classical Be stars.  They devised a method independent
of Lucy's (2002) transition 
probability method to solve the equations of statistical equilibrium.   It is similar in many ways,
except that the photon absorption and re-emission mechanisms are uncorrelated, allowing them
to dispense with Lucy's macro atoms, along with their associated internal transitions
and Monte Carlo transition probabilities.
Their models show that the optically thick regions of the disk are similar to Young Stellar Object (YSO)
disks and the optically thin outer parts are like stellar winds.
Carciofi \& Bjorkman (2008) build on their previous work and solve the steady state nonisothermal viscous diffusion
and vertical hydrostatic equilibrium of Keplerian disks.
Their solution departs significantly from the analytic isothermal density, affecting the emergent spectrum.

\subsubsection{Photoionization}

Several authors describe algorithms for calculating photoionization, e.g., Och et al. (1998), Wood \& Loeb (2000),
Ciardi et al. (2001), Maselli, Ferrara \& Ciardi (2003), Ercolana et al. (2003), Wood, Mathis
\& Ercolana (2004), Ercolana et al. (2008), and Cantalupo \& Porciana (2011).
Some particular features of these codes are Wood et al.'s (2004) use of
photon packets vs energy packets to more easily match the notation of the recombination coefficients;
the x-ray extension to the MOCASSIN code to allow computation detailed high-resolution spectra
(Ercolano et al. 2008);
and photoionization on adaptive mesh refinement grids (Cantalupo \& Porciani 2011).

These have been applied to the study of escape of ionizing radiation from
high-redshift galaxies (Wood \& Loeb 2000), cosmological reionization around the first
stars (Ciardi et al. 2001), modeling the diffuse ionized gas in the Milky Way and other galaxies
(Wood \& Mathis 2004,  photoevaporating planetary disks (Ercolano \& Owen 2010), H II regions 
(Ercolano, Wesson \& Bastian 2010 and references therein), and planetary nebulae (Ercolano et al. 2004 and
references therein), to name a few.

\subsubsection{Chemistry}

The combinations of dust radiative transfer (Section 4.1) and line radiative transfer (Section 4.6.1) can be used to
study the chemistry in clouds.  Jorgensen et al. (2006) iterate on the dust temperature
and molecular line calculations to determine where molecules freeze-out in protostellar envelopes.
Spaans (1996) includes a chemical network of 44 species to study the effects of clumpiness.
Bruderer et al. (2009a,b, 2010) demonstrate chemical modeling of Young Stellar Objects
 in a 3-part series.   They pre-calculate a grid of chemical composition as a function of time,
 for a given gas density, temperature, far-UV irradiation and X-ray flux. The local far-UV flux
 is calculated by a Monte Carlo radiative transfer code, which includes scattering and temperature
 calculation. The use of the pre-calculated chemical grid speeds up calculations by several orders of magnitude.

\section{Summary}

The Monte Carlo method for radiative transfer (MCRT) is complementary to the traditional formal
methods.  While those excel in 1-D, at high optical-depths, incorporating many gas lines and 
computing detailed spectra, MCRT excels with 3-D geometries, non-LTE gas processes, anisotropic radiation fields and
scattering functions, complex velocity fields, and polarization calculations.   
Thus MCRT is a great tool to add to the set of well-developed methods for radiative transfer.
In fact, it is a necessary tool to interpret the ever-increasing sophistication of our new
observatories.

\section*{Acknowledgements}
I thank the editors for inviting me to write this article in celebration of the Chandra Centennial.  It is an honour to be asked.

\label{lastpage}
\end{document}